# The gmwm R package: a comprehensive tool for time series analysis from state-space models to robustness

**James Balamuta**
Univeristy of Illinois

**Stéphane Guerrier**
Univeristy of Illinois

**Roberto Molinari**
University of Geneva

**Wenchao Yang**
Univeristy of Illinois

#### Abstract

The **gmwm** R package for inference on time series models is mainly based on the quantity called wavelet variance which is derived from a wavelet decomposition of a time series. This quantity provides a means to summarize and graphically represent the features of time series in order to identify possible models. Moreover, it is used as a moment condition for model estimation through the generalized method of wavelet moments. Based on the latter method, this package not only provides an alternative method to estimate classical ARMA models but also delivers a general framework for the robust estimation of many time series models as well as a quick and efficient estimation of many linear state-space models.

## 1. Introduction

Time series analysis in R can be carried out with a wide range of tools and packages. However, there are many practical limitations to the methods which are currently implemented when it comes to specific requirements such as, for example, a user-friendly estimation of certain state-space models, robust inference for time series models and estimation on large datasets. The new **gmwm** R package helps to overcome some of these limitations by implementing the Generalized Method of Wavelet Moments (GMWM), which was proposed by Guerrier, Skaloud, Stebler, and Victoria-Feser (2013), along with its relative inference tools. This method uses a quantity called the Wavelet Variance (WV), which is the variance of the wavelet coefficients that are issued from a wavelet decomposition of a time series (see for example Percival and Walden 2006). The WV is a widely used quantity in different fields such as geology and aerospace engineering since it helps to decompose and interpret the variance of a time series across different "scales" and represents a good statistic to summarize the "information" of time series which respect certain properties (e.g. intrinsically stationary). The GMWM uses this quantity as an auxiliary parameter in a minimum distance estimator setting allowing it to estimate a wide range of intrinsically second-order stationary models in a numerically stable and computationally efficient manner.

In this paper, we will focus primarily on some specific features of the **gmwm** package, highlighting other important features in the process. More specifically, this package delivers two main advantages for univariate[1] time series model estimation compared to currently available packages: (i) easy-to-use, computationally efficient and numerically stable estimation of linear state-space models and (ii) robust estimation and inference for the whole range of time series models available for estimation in the package. Moreover, for these two aspects, a related model selection procedure is available based on the Wavelet Information Criterion (WIC) (see Guerrier, Molinari, and Skaloud 2015). For this reason we will list the main tools available in R for these two particular goals whereas we refer the reader to Shumway and Stoffer (2013) and the R website[2] for a more general overview of all available packages and functions for time series analysis.

Firstly, when dealing with parameter estimation of state-space models there is a relative abundance of options in R for which we provide a non-exhaustive summary. The **KFKSDS** package, for example, allows to treat state-space models via Kalman-filtering as does the **dlm** package for Gaussian linear state-space models. The **bsts** package uses Markov-Chain Monte-Carlo to simulate from the posterior distribution of Bayesian structural time series models which provides estimates for the coefficients representing the states in the desired model. The **dse** package represents a general framework for linear and multivariate time series estimation with general AutoRegressive Moving Average (ARMA) representations and functions allowing for the conversion of these models to state-space model representations. An even more general and computationally efficient framework is represented by **KFK** where a fast implementation of the Kalman filter is available to estimate state-space models also for large datasets while the **KFAS** package also makes forecasting tools available. The **dlm** package represents one of the main available packages when it comes to maximum-likelihood estimation of Gaussian linear state-space models which includes Bayesian analysis tools as well as time series smoothing options. Along these lines we also find the **MARSS** package which allows for the estimation of multivariate first-order autoregressive state-space models with Gaussian errors via the Expectation-Maximization algorithm, with the additional possibility of constraining the estimation procedure and adding covariates. Finally, the **pomp** package extends the estimation possibilities to non-Gaussian and non-linear state-space models by using the framework of partially observed Markov processes.

On the other hand, as far as robustness is concerned, the availability of R functions that allow robust estimation and inference for time series models is almost nonexistent for classic models such as ARMA, without mentioning state-space models. The only package which directly deals with robustness for time series is **robfilter** which makes available a series of robust filters to smooth the observed time series. Another package which includes robustness for time series analysis is **robcor** in which the function `robacf()` delivers a robust equivalent of the function `acf()` to compute the empirical autocovariance and autocorrelation function. However, none of these are able to deal directly with robust parametric estimation and inference for time series models. An interesting option that is available for this purpose is the **quantreg** package in which the function `dynrq()` allows to perform robust-type estimation of autoregressive models (being based on quantiles). A "naive" approach resembling the latter method would be to use the robust regression functions such as `rlm()` or `lmrob()`, in the **MASS** and **robustbase** packages respectively, to also robustly estimate autoregressive models. These autoregressive models can then eventually be used as auxiliary models to estimate more complex models via indirect inference (see Gourieroux, Monfort, and Renault 1993) but this can come at the price of computational inefficiency and numerical convergence problems when the models are complex.

---

[1] The extension to multivariate time series is a current research topic.
[2] https://cran.r-project.org/web/views/TimeSeries.html

As a final note, as highlighted above there are different solutions which are either directly available or can be obtained using different tools in R. Nevertheless, even when solutions are available, these may not necessarily be numerically stable or computationally efficient as stated above. For many applications, such as economics, the time series are roughly of length $T = 1,000$ or smaller and, if discarding the need for robust analysis, these problems can easily be treated using currently available tools. However, there are many other applications where the observed time series can easily be much longer (sometimes in the order of $500,000$ or 1 million) and a comprehensive estimation and inference procedure for these cases is hardly feasible by using the available tools in R

Considering the above, the **gmwm** package adds to the state-space model estimation tools by providing numerically stable and computationally efficient estimations also for large datasets while delivering the first readily available software for the robust estimation of a wide range of instrinsically second-order stationary (state-space) models based on the Robust GMWM (RGMWM) proposed in Guerrier and Molinari (2016a). All this is available along with a series of inference and model selection tools which allow to have a general purpose package for time series analysis. With this in mind, Section 2 briefly introduces the syntax used in the package by focusing on time series model simulation, thereby listing the time series models available for simulation and estimation purposes. Section 3 discusses the tools the package makes available for computing and representing the WV of a time series while in Section 4 we present the estimation and inference tools for different time series models from ARMA to state-space models where their corresponding robust framework is described. Finally, Section 5 briefly discusses the computational efficiency of the new package and Section 6 concludes by providing a list of upcoming and future features with which the package will be updated.

## 2. Time Series Models

In this section we briefly list, describe and provide the syntax for the models available for estimation and simulation purposes within the **gmwm** package. Under some minor constraints, all these models can then be combined into a specific class of linear state-space models which can be represented as a sum of underlying models. Having stated this, some of the basic models available in the **gmwm** package to simulate from (and to estimate) are the following:

- White Noise (`WN()`);
- Quantization Noise (`QN()`);
- Random Walk (`RW()`);

- Drift (`DR()`);

- AR(1): First-order autoregressive process (`AR1()`);

- ARIMA: Integrated Autoregressive Moving Average process (`ARIMA()`);

- SARIMA: Seasonal ARIMA process (`SARIMA()`).

The expressions in brackets in the above list represent the syntax used in the **gmwm** package to specify the model. The only model which may be less known is the quantization noise which is a process that is often used in engineering fields and can be described in layperson terms as being a good approximation of a rounding error. The brackets for the syntax of each model are left to specify parameter values for simulation purposes or to specify starting values for estimation purposes. The code below shows how these model specifications can be used for simulations based on the built-in function `gen.gts()` which allows to generate samples from all these models.

```r
# Set seed for reproducibility
set.seed(1337)

# Number of observations
n = 1e4

# Generate a White Noise Process
wn = gen.gts(WN(sigma2 = 1), n)

# Generate a Quantization Noise
qn = gen.gts(QN(q2 = .5), n)

# Generate a Random Walk
rw = gen.gts(RW(gamma2 = .75), n)

# Generate a Drift
dr = gen.gts(DR(omega = 0.10), n)

# Generate an AR(1)
ar1 = gen.gts(AR1(phi = .9, sigma2 = .1), n)

# Generate an MA(1)
ma1 = gen.gts(MA1(theta = .3, sigma2 = .5), n)

# Generate an ARMA(1,1)
arma11 = gen.gts(ARMA11(phi = .9, theta = .2, sigma2 = 1), n)

# Generate an SARIMA(1,0,1)x(2,1,1)[12] process
sarima = gen.gts(SARIMA(ar = 0.3, i = 0, ma = -0.27,
                    sar = c(-0.12, -0.2), si = 1, sma = -0.9,
                    sigma2 = 1.5, s = 12), n)
```

The **gmwm** package therefore allows to easily simulate from a wide variety of models, including SARIMA models, but does not limit itself to these basic models. Indeed, under some restrictions highlighted in Section 4, these models can be combined in different ways to deliver many state-space (*latent*) models which can be represented by the sum of basic models. The construction of such linear state-space models is very simple with the **gmwm** package allowing it to be considerably user-friendly. In fact, to specify that a model is a combination of different models, all that is needed is to use the "+" symbol between them and, supposing that different `AR1()` processes are present in the state-space model, the syntax to insert "`k`" of these models in a state-space model is `k*AR1()`. So, for example, the sum of three AR(1) models, a random walk and a white noise process can be specified as: `3*AR1()+RW()+WN()`. A function specifically provided to generate and represent these models is the `gen.lts()` which, while simulating from these models, also gives the option to plot a breakdown of the underlying processes by applying the `plot()` function on the result of `gen.lts()` (see Section 4 for an example of this feature). It must be noted that this particular function is designed specifically to highlight how these state-space models are obtained and that the `gen.gts()` function should instead be used if the sole purpose is to simulate time series since it is computationally faster.

## 3. Modeling based on the Wavelet Variance

As highlighted in the introduction, the WV is a very useful quantity for the analysis of time series and provides the auxiliary parameter for the GMWM. Therefore in this section we discuss how this quantity is estimated in the package and underline how it can be used for modelling purposes, consequently inviting the reader to go directly to Section 4 if their interest lies mainly in model estimation and inference. To introduce the WV, which we denote as $\nu_j^2$, let $(Y_t)$ represent a time series and let $(W_{j,t})$ denote the wavelet coefficient process issued from the a Haar wavelet decomposition of $(Y_t)$ at the $j^{th}$ scale (Percival and Walden 2006). Then the WV at scale $j$ can be defined as $\text{Var}(W_{j,t})$ (i.e. the variance of the wavelet coefficients) and an unbiased estimator for this quantity was proposed by Percival (1995) and is given by

$$\hat{\nu}_j^2 = \frac{1}{M_j} \sum_{t=1}^{M_j} W_{j,t}^2,$$

where $M_j$ represents the number of wavelet coefficients at the considered scale. An M-estimator which generalizes this estimator was proposed by Guerrier and Molinari (2016c) which can be made robust by selecting a bounded estimating function which, for the **gmwm** package, consists in the Tukey biweight function (Beaton and Tukey 1974). The function that is used for the estimation of these quantities is called `wvar()`, where the option `robust=T` allows to make use of the robust estimator of WV. The output of this function is a vector of estimated WV of length $J = \lfloor \log_2(T) \rfloor$, with $T$ being the length of the observed time series and $\lfloor x \rfloor$ representing the largest integer smaller than $x$, along with the element-wise confidence intervals.

The WV is widely used for non-parametric analysis of time series, for example to interpret natural or economic phenomena across different time-scales (see Percival and Walden 2006). In some engineering fields, such as navigation, a commonly used quantity is the Allan Variance (AV) which is directly related to the Haar WV (i.e. AV $\equiv$ 2WV) and supports practitioners who want to identify certain behaviors in the observed time series. Indeed, as mentioned earlier, the WV well summarizes the information contained in the spectral density and, for example, a plot of the logarithmic transform of WV versus its scales can help to understand the kind of models which can be generating the observed process since different models have different linear (or non-linear) behaviors when represented in this manner. Therefore, this visualization tool represents another helpful option to understand what kind of models could explain the time series, just like the basic AutoCorrelation Function (ACF) and Partial Auto-Correlation Function (PACF) plots can help understand the order of certain ARMA models for example.

Considering the usefulness of these graphical representations, the **gmwm** package allows to easily generate these figures by applying the `plot()` function to the object issued from the `wvar()` function. An additional function which allows to graphically compare the WV issued from different time series or to compare the standard and robust WV on the same time series is `compare.wvar()`. This is a very useful tool if one is interested in understanding if different time series can be explained by similar models or if the time series suffers from outliers or other forms of contamination. To provide some examples, the code below estimates the WV from some of the time series simulated in Section 2 and produces the plots in Figure 1[3].

```
# Compute WV
wv.wn     = wvar(wn)
wv.qn     = wvar(qn)
wv.rw     = wvar(rw)
wv.dr     = wvar(dr)
wv.ar1    = wvar(ar1)
wv.sarima = wvar(sarima)

# Plot WV
compare.wvar(wv.wn, wv.qn,
             split = F, auto.label.wvar = F,
             legend.label = c("WN","QN"))
compare.wvar(wv.rw, wv.dr,
             split = F, auto.label.wvar = F,
             legend.label = c("RW","DR"))
plot(wv.ar1)
plot(wv.sarima)
```

A first detail which helps to interpret these graphs is that for stationary processes we generally have that the WV decreases at the larger scales while it increases for non-stationary ones. For example, a WV plot which decreases linearly with a certain slope indicates that the process could be white noise while a linearly increasing WV could indicate a random walk or a drift (depending on the slope of the WV). A plot of the WV which does not have a clear linear behavior could instead indicate the presence of a more complex process such as ARMA processes or a combination of different models. Indeed, models like white noise and quantization noise appear as negative linear functions of the scales of decomposition (which are denoted as $\tau$ in the plots) with different slopes (i.e. steeper for the quantization noise) while drift and random walk models appear as positive linear functions where the drift has a steeper slope. The plots of the AR(1) and SARIMA processes show how these models don't exactly have a linear behaviour in terms of WV. As can be observed, these plots by default give confidence intervals for the estimated WV which allow to understand if, for example, there is a significant difference between the standard and robust estimators of WV, thereby indicating that there could be contamination in the data. As an example, let us take the production index data made available, among others, in the **astsa** package and named `prodn`. Shumway and Stoffer (2013) suggest a SARIMA$(2, 1, 0) \times (0, 1, 3)_{12}$ as a good model for this data so, taking the appropriate differences, the output of the following commands is given in Figure 2.

---

[3] All the figures in this paper are enhanced versions of the **gmwm** package plots.

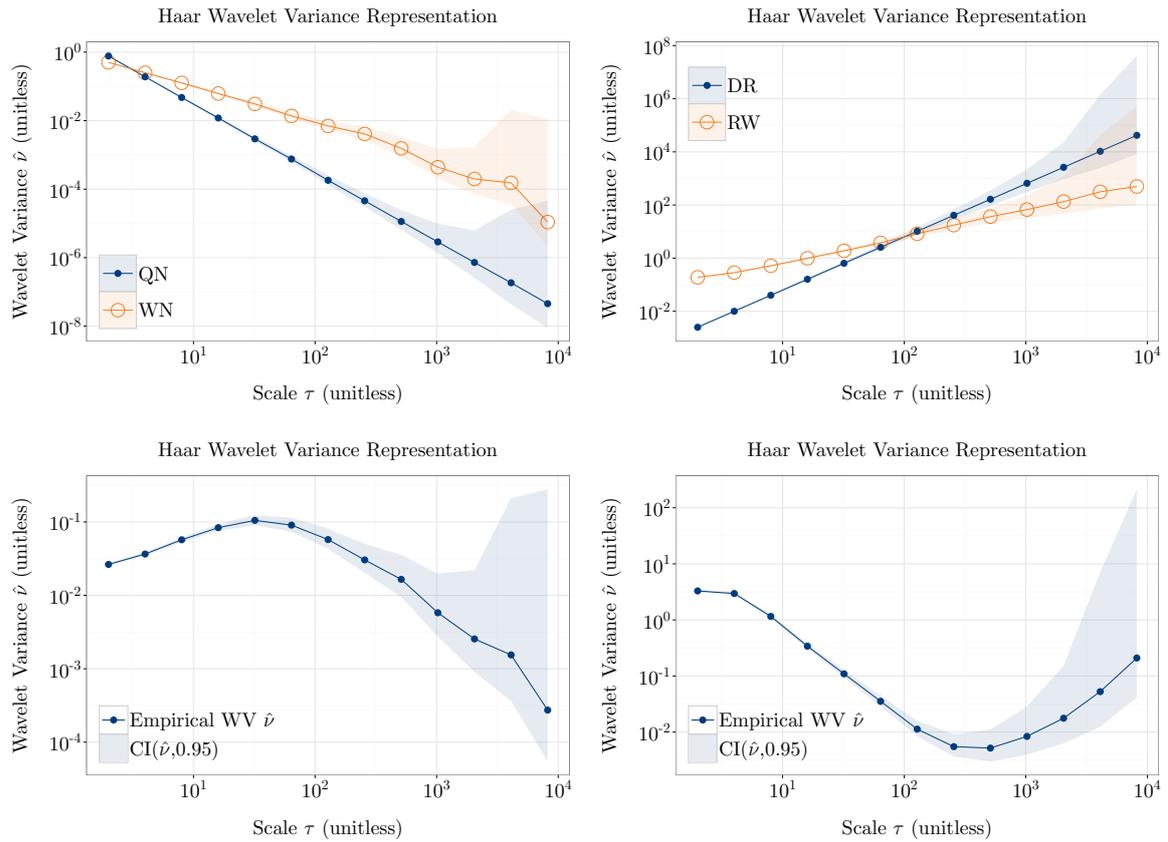

Figure 1: Top-left: Estimated WV for white noise (orange line) and estimated WV for quantization noise (blue line) with confidence intervals (shaded areas). Top-right: Estimated WV for random walk (orange line) and estimated WV for drift (blue line) with confidence intervals (shaded areas). Bottom-left: Estimated WV for an AR(1) process with confidence intervals (shaded areas). Bottom-right: Estimated WV for an SARIMA$(1,0,1) \times (2,1,1)_{12}$ process process with confidence intervals (shaded areas).

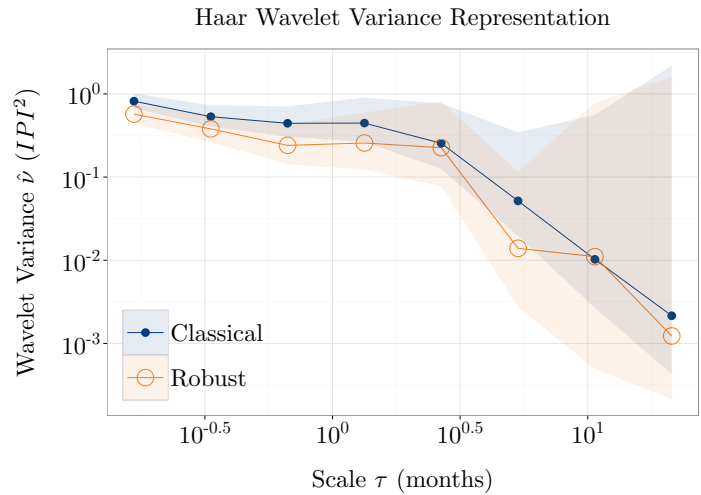

Figure 2: Comparison between standard and robust WV for the `prodn` data.

```
# Take Non-seasonal and seasonal differences
prodn_diff = diff(diff(prodn), 12)

# Compute standard and robust WV
wv.prodn_diff = wvar(prodn_diff)
wv.prodn_diff.r = wvar(prodn_diff, robust = TRUE)

# Compare standard and robust WV
compare.wvar(wv.prodn_diff, wv.prodn_diff.r, split = FALSE)
```

It can be seen how even though there are slight differences between the standard and robust WV estimates, the confidence intervals show that these differences do not appear to be significant and therefore a classical analysis appears to be appropriate for this data.

# 4. Model Estimation and Inference

The GMWM was initially developed for the estimation of so-called *latent* or *composite* processes in the context of sensor calibration. This estimator is defined as follows

$$\hat{\boldsymbol{\theta}} = \underset{\boldsymbol{\theta} \in \boldsymbol{\Theta}}{\operatorname{argmin}} \, (\hat{\boldsymbol{\nu}} - \boldsymbol{\nu}(\boldsymbol{\theta}))^T \, \boldsymbol{\Omega} \, (\hat{\boldsymbol{\nu}} - \boldsymbol{\nu}(\boldsymbol{\theta})) \qquad (1)$$

where $\boldsymbol{\theta}$ represents the time series model parameter vector that we intend to estimate belonging to the compact set $\boldsymbol{\Theta}$; $\hat{\boldsymbol{\nu}} = [\nu_j^2]_{j=1,\ldots,J}$ represents the vector of estimated WV; $\boldsymbol{\nu}(\boldsymbol{\theta}) = [\nu_j^2(\boldsymbol{\theta})]_{j=1,\ldots,J}$ represents the WV implied by the model; $\boldsymbol{\Omega}$ is a positive definite weighting matrix chosen in a suitable manner (see Guerrier *et al.* 2013). The general estimation function which implements this estimator is called `gmwm()` for which different options can be specified such as, for example, the method to estimate the weighting matrix $\boldsymbol{\Omega}$. However, the main arguments for this function are of course the observed time series as well as the model that we intend to estimate (see Section 2).

## 4.1. The class of SARIMA models

The class of SARIMA models includes many of the basic models that can be estimated within the **gmwm** package, with a few exceptions such as quantization noise. This is possible based on the results in Zhang (2008) where analytic expressions for $\boldsymbol{\nu}(\boldsymbol{\theta})$ can be obtained for all the models listed in Section 2. Given this, the package therefore delivers an additional tool to estimate SARIMA models in R, with a possible advantage residing in the fact that it can easily estimate these models even for considerably large sample sizes. To see how these models can be estimated with the `gmwm()` function, let us take the `prodn` data whose WV we analysed at the end of Section 3. As mentioned earlier, Shumway and Stoffer (2013) suggest a SARIMA$(2,1,0) \times (0,1,3)_{12}$ model for this data so let us estimate this using also the standard `arima()` function for Maximum Likelihood Estimation (MLE) for comparison. The code for these two procedures can be found below.

```
# MLE
mle.fit = arima(prodn,
                order = c(2,1,0),
                seasonal = list(order = c(0,1,3), period = 12),
                include.mean = FALSE)

# GMWM
gmwm.fit = gmwm(SARIMA(ar = 2, i = 1, ma = 0, sar = 0, si = 1, sma = 3),
                data = prodn)

# View GMWM Fit
plot(gmwm.fit)
```

The SARIMA model is specified within the `gmwm()` function by using the syntax `SARIMA()` in which the orders are defined and its result is stored in an object called `gmwm.fit`. The results of the `arima()` and `gmwm()` estimation functions, along with their confidence intervals, are given in Table 1.

|  | GMWM | | MLE | |
|---|---|---|---|---|
|  | Estimate | 95% CI | Estimate | 95% CI |
| $\phi_1$ | $2.972 \cdot 10^{-1}$ | $[\ 1.875 \cdot 10^{-1},\ \ 3.913 \cdot 10^{-1}]$ | $3.038 \cdot 10^{-1}$ | $[\ 1.920 \cdot 10^{-1},\ \ 4.120 \cdot 10^{-1}]$ |
| $\phi_2$ | $1.760 \cdot 10^{-1}$ | $[\ 6.236 \cdot 10^{-2},\ \ 2.880 \cdot 10^{-1}]$ | $1.077 \cdot 10^{-1}$ | $[\ 1.294 \cdot 10^{-2},\ \ 2.309 \cdot 10^{-1}]$ |
| $\Theta_1$ | $-8.872 \cdot 10^{-1}$ | $[-1.045 \cdot 10^{0}\ ,-6.207 \cdot 10^{-1}]$ | $-7.393 \cdot 10^{-1}$ | $[-8.254 \cdot 10^{-1},-6.373 \cdot 10^{-1}]$ |
| $\Theta_2$ | $-2.764 \cdot 10^{-2}$ | $[-5.344 \cdot 10^{-1},\ \ 6.637 \cdot 10^{-1}]$ | $-1.445 \cdot 10^{-1}$ | $[-2.983 \cdot 10^{-1},-3.477 \cdot 10^{-2}]$ |
| $\Theta_3$ | $2.864 \cdot 10^{-1}$ | $[-3.386 \cdot 10^{-1},\ \ 8.071 \cdot 10^{-1}]$ | $2.815 \cdot 10^{-1}$ | $[\ 1.790 \cdot 10^{-1},\ \ 3.883 \cdot 10^{-1}]$ |
| $\sigma^2$ | $1.154 \cdot 10^{0}$ | $[\ 9.060 \cdot 10^{-1},\ \ 1.238 \cdot 10^{0}\ ]$ | $1.312 \cdot 10^{0}$ | $[\ 1.100 \cdot 10^{0}\ \ ,\ \ 1.471 \cdot 10^{0}\ ]$ |

Table 1: MLE and GMWM estimated parameters for the SARIMA model on the `prodn` data. $\phi_i$ represents the $i^{th}$ autoregressive parameter, $\Theta_i$ represents the $i^{th}$ seasonal moving-average and $\sigma^2$ represents the innovation variance.

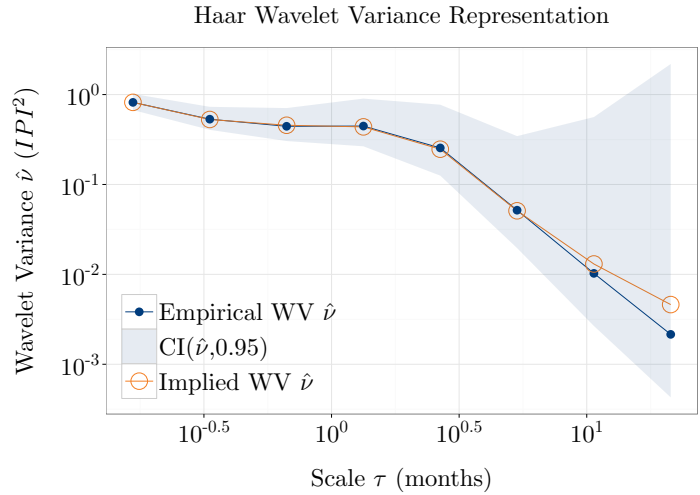

Figure 3: Estimated WV from the `prodn` data (blue line) with confidence intervals (shaded area) and WV implied by estimated SARIMA model (orange line)

There appear to be no significant differences between the two methods, although the last two seasonal moving-average parameters do not appear to be significantly different from zero for the GMWM. Once the model has been estimated and saved into an object (i.e. `gmwm.fit` in the above code), it is possible to make a visual assessment of how well the model fits the observed time series by comparing the estimated WV with the WV implied by the estimated parameters simply by applying the function `plot()` to the estimated model. This can be seen in Figure 3 where the orange line, representing the WV implied by the estimated SARIMA model, closely follows the WV estimated directly on the `prodn` data (blue line) and lies within its confidence intervals (shaded area).

### 4.2. Linear State-Space Models

As mentioned in Section 2, the **gmwm** package is able to estimate all those linear state-space models which can be represented as the sum of different underlying processes. These models extend from a wide variety of linear state-space models to structural time series models. An example of these models is given by the process $(Y_t)$ defined as follows

$$\begin{aligned} X_t &= X_{t-1} + \delta + \epsilon_t, \ \epsilon_t \overset{iid}{\sim} \mathcal{N}(0, \upsilon^2) \\ Y_t &= X_t + U_t, \ U_t \overset{iid}{\sim} \mathcal{N}(0, \sigma^2) \end{aligned}$$

where $\delta$ represents a drift. This state-space model, also known as a local-linear trend model, is often used in ecology, for example, to describe population dynamics with measurement errors and basically corresponds to the sum of a white noise, a random walk and a drift (an example of how this model is estimated within the package is provided further on). The reason why the GMWM manages to estimate these kind of models in a computationally efficient manner is because their WV is simply the result of the sum of individual WV of the underlying models whose analytic expressions are known (see Section 4.1) and the number of scales (i.e. the number of auxiliary parameters) is always reasonable even for large sample sizes. Therefore many models can be summed together to deliver different types of latent models. There are however some restrictions on certain model specifications due to parameter identifiability issues as underlined in Guerrier and Molinari (2016b). For example, we can have a `WN()`, a `QN()`, a `RW()` and a `DR()` only once in a state-space model whereas an `ARMA()` or an `AR1()` can possibly be included as many times as desired. It must be specified however that it is not certain that a sum of ARMA models is identifiable while a sum of AR(1) models, which well approximates many processes, is generally identifiable.

Let us now give a simple example to illustrate how the **gmwm** package estimates these types of models and therefore consider the process $(Y_t)$ defined as follows

$$\begin{aligned} X_t &= \phi_1 X_{t-1} + \epsilon_t, \ \epsilon_t \overset{iid}{\sim} \mathcal{N}(0, \upsilon^2), \\ W_t &= \phi_2 W_{t-1} + \varepsilon_t, \ \varepsilon_t \overset{iid}{\sim} \mathcal{N}(0, \sigma^2), \\ Y_t &= X_t + W_t + Z_t \end{aligned} \qquad (2)$$

where $Z_t = t\delta$ is a drift process. The process $(Y_t)$ is hence the result of the sum of two first-order autoregressive processes and a drift process. This is another example of a state-space model where the states are represented by the unobserved processes $(X_t)$, $(W_t)$ and $(Z_t)$. For this model, let us moreover define the parameter values as follows: $\phi_1 = 0.9$, $\upsilon^2 = 1$, $\phi_2 = 0.1$, $\sigma^2 = 4$ and $\delta = 0.01$. The code below simulates from this model using the `gen.lts()` function mentioned in Section 2, consequently plotting the breakdown of this state-space model (left plot of Figure 4) and then estimating its parameters using the `gmwm()` function.

```r
# Length of time series
n = 1e3

# Define model
true.model = AR1(0.9,1) + AR1(0.1,4) + DR(0.01)

# Simulate and plot breakdown of time series
set.seed(1337)
sim.ts = gen.lts(true.model, n)
plot(sim.ts)
```

```
# Estimate model
model = 2*AR1() + DR()
fit = gmwm(model, sim.ts)

# Visualize WV of estimated model
plot(fit, process.decomp = TRUE)

# Model inference
inference = summary(fit, inference = TRUE, bs.gof = TRUE)
```

As highlighted Section 4.1, the object containing the result of the estimation procedure (i.e. `fit` in the example code above) can be used within the `plot()` function and, in particular, if specifying the option "`process.decomp = TRUE`" within this function, the user will obtain a plot similar to the one presented in the right plot of Figure 4. The blue line represents the estimated WV while the orange one represents the WV implied by the estimated model. The other lines are the result of specifying the option "`process.decomp = TRUE`" and show the WV implied by the estimated parameters for the individual models composing the overall sum. It can be seen how the individual WV help to "support" the overall implied WV in order to fit the estimated WV and, in this case, the fitted model appears to be a good one since it closely follows the estimated WV and lies within its confidence intervals (shaded area). This is confirmed by the Goodness-of-Fit (GoF) test contained in the `inference` object generated by executing the command `summary(fit, inference = TRUE, bs.gof = TRUE)` whose output can be found below.

```
Model Information:
       Estimates       CI Low     CI High           SE
AR1    0.8808235 0.800135885  0.94028587  0.042602569
SIGMA2 0.7454827 0.216068102  1.46524184  0.379721853
AR1    0.2025861 0.075190570  0.30140290  0.068763666
SIGMA2 4.5995931 3.659867741  5.42981374  0.538025380
DR     0.0095902 0.005945415  0.01409644  0.002477737

Objective Function: 0.0012

Bootstrapped Goodness of Fit:
Test Statistic: 0
P-Value: 0.97  CI: (0.93, 1)

To replicate the results, use seed: 1337
```

It can be seen how the result of this procedure is a table containing the estimated parameters, their confidence intervals and, as already mentioned, the details on the GoF test which is based on the Sargan-Hansen test (also known as "J-test", see Sargan 1958). For the latter, it is possible to specify the option `bs.gof = T` which means that the distribution of the test-statistic (i.e. the value of the GMWM objective function at the estimated parameter) is approximated by parametric bootstrap[4].

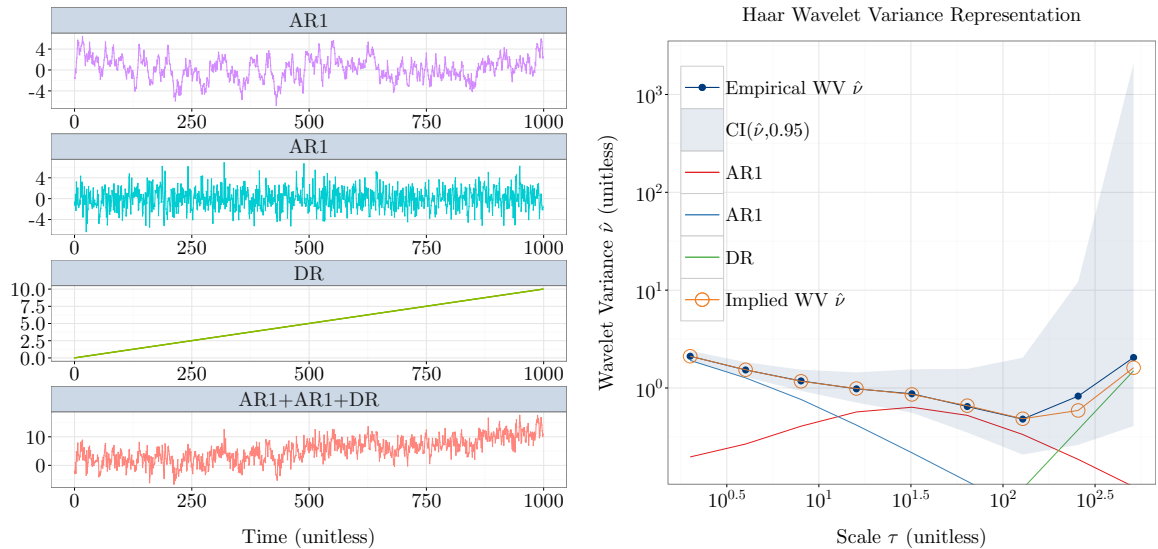

Figure 4: Left: Breakdown of underlying processes and resulting latent process (bottom) from model specified in (2). Right: WV implied by the fitted model compared to the estimated WV. Additional lines show the WV implied by the parameters of each of the underlying models.

Let us now use the **gmwm** package on a real dataset containing measurements on the annual flow of the Nile river at Ashwan between 1871 and 1970 available in R within the **datasets** package. Petris and Petrone (2011) suggest two structural models for this dataset:

- **Local level model**: this model is given by the sum of a white noise model and a random walk;

- **Local linear trend**: this model is given by the sum of a local level model and a drift (mentioned as an example at the beginning of this section).

These models can both be estimated with the **gmwm** package as shown in the code further on. The objects `loc.level` and `loc.trend` contain the estimations of the local level and local linear trend models respectively. In addition to the estimation of these state-space (structural) models, the **gmwm** package allows to perform a GoF test, as seen earlier in the simulated example, as well as to select the "best" model based on the mentioned the WIC. The latter criterion tells us in very general terms how well the estimated model can predict the values of the WV issued from another realization of the same process and is defined as follows:

$$\text{WIC} = \mathbb{E}\left[\mathbb{E}_0\left[\left(\hat{\boldsymbol{\nu}}^0 - \boldsymbol{\nu}(\hat{\boldsymbol{\theta}})\right)^T \boldsymbol{\Omega} \left(\hat{\boldsymbol{\nu}}^0 - \boldsymbol{\nu}(\hat{\boldsymbol{\theta}})\right)\right]\right] \tag{3}$$

where $\mathbb{E}_0[\cdot]$ and $\hat{\boldsymbol{\nu}}^0$ denote respectively the expectation and the estimated WV based on

---

[4]NB: The GoF test statistic and p-value in the output are rounded to two decimal places.

another realization of the same process. Its corresponding estimator is given by

$$\widehat{\text{WIC}} = \left(\hat{\boldsymbol{\nu}} - \boldsymbol{\nu}(\hat{\boldsymbol{\theta}})\right)^T \boldsymbol{\Omega} \left(\hat{\boldsymbol{\nu}} - \boldsymbol{\nu}(\hat{\boldsymbol{\theta}})\right) + 2\operatorname{tr}\left\{\widehat{\text{cov}}\left[\hat{\boldsymbol{\nu}}, \boldsymbol{\nu}(\hat{\boldsymbol{\theta}})\right] \boldsymbol{\Omega}^T\right\} \tag{4}$$

where $\widehat{\text{cov}}(\cdot)$ denotes a consistent estimator of the covariance. The first term in the estimator corresponds to the value of the objective function of the GMWM at the estimated parameter $\hat{\boldsymbol{\theta}}$ while the second term is sometimes referred to as "optimism". Indeed, the first term is expected to decrease as parameters are added while the second is expected to increase in these cases, thereby providing some kind of penalty for overfitting. This estimator is computed within the function `rank.models()` where the user can provide a set of *nested* candidate models, as well as the dataset, to finally obtain a list of models ranked in ascending order according to the WIC[5]. Moreover, it is possible to specify how to compute $\widehat{\text{cov}}(\cdot)$ in the second term, which is either through parametric bootstrap (see Guerrier *et al.* 2015) or through an analytic form (see Zhang and Guerrier 2015). Since the aim is to minimize this quantity, the first model in the list can be considered as the one which best predicts the WV and therefore, generally speaking, the time series itself.

```
# Retrieve Nile dataset
nile = datasets::Nile

# Estimate the models
loc.level = gmwm(WN() + RW(), nile)
loc.trend = gmwm(WN() + RW() + DR(), nile)

# Goodness-of-fit of the models
> summary(loc.level, inference = TRUE, bs.gof = TRUE)

Model Information:
    Estimates    CI Low    CI High       SE
WN  13611.69   9133.942  17598.004  2572.892
RW   2095.55    167.214   3787.207  1100.400

Objective Function: 0.0288

Bootstrapped Goodness of Fit:
Test Statistic: 0.03
P-Value: 0.24   CI: (0.16, 0.33)

To replicate the results, use seed: 1337

> summary(loc.trend, inference = TRUE, bs.gof = TRUE)

Model Information:
      Estimates         CI Low        CI High            SE
```

---

[5]The models for which the WIC can be computed through its analytical form are the models (and combination thereof): `WN()`, `QN()`, `RW()`, `DR()`, `AR1()`, `MA1()` and `ARMA11()`.

```
WN 1.361398e+04 9135.6649579 1.759593e+04 2.571737e+03
RW 2.092888e+03  183.5483823 3.767013e+03 1.089296e+03
DR 4.928475e-01    0.4897141 4.946456e-01 1.499095e-03

Objective Function: 0.0289

Bootstrapped Goodness of Fit:
Test Statistic: 0.03
P-Value: 0.26  CI: (0.18, 0.35)

To replicate the results, use seed: 1337

# Select model
WIC = rank.models(RW() + WN(), RW() + WN() + DR(), data = nile)
> WIC

The model ranking is given as:
            Obj Fun Optimism Criterion
1. RW WN     0.0288   0.9289    0.9577
2. RW WN DR  0.0289   0.9314    0.9602

plot(WIC)
```

As can be observed, the two models appear to be close in terms of the WIC which is given in the last column (the first two columns are the first and second term of (4) respectively). However, following the rule of picking the model with the lowest WIC, this would be the local-level model and, if applying the function `plot()` to the object issued from the `rank.models()` function (i.e. `plot(WIC)`), it is possible to assess how well the selected model fits the observed WV. In this case Figure 5 shows that the WV implied by the selected model fits the observed WV well at the first scales and less so at the last, remaining however strictly within the confidence intervals.

### 4.3. Robust Inference for Time Series Models

There has been a large amount of research dedicated to providing sound statistical methods to deliver robust estimation and inference for time series models (see Maronna, Martin, and Yohai 2006). However, as mentioned in the introduction, the implementation of these methods has been scarce (if not nonexistent) and researchers and practitioners have not been able to make use of these approaches which are often complicated to implement and numerically unstable. The **gmwm** package offers a first stable solution to this gap in the available software by simply replacing the estimated WV $\hat{\nu}$ in (1) with the robust M-estimator proposed by Guerrier and Molinari (2016c) which has well defined asymptotic properties as well as better finite sample performance compared to existing estimators. To perform a robust estimation of time series model parameters within the package, all that is needed is to specify the option `robust = TRUE` within the `gmwm()` function.

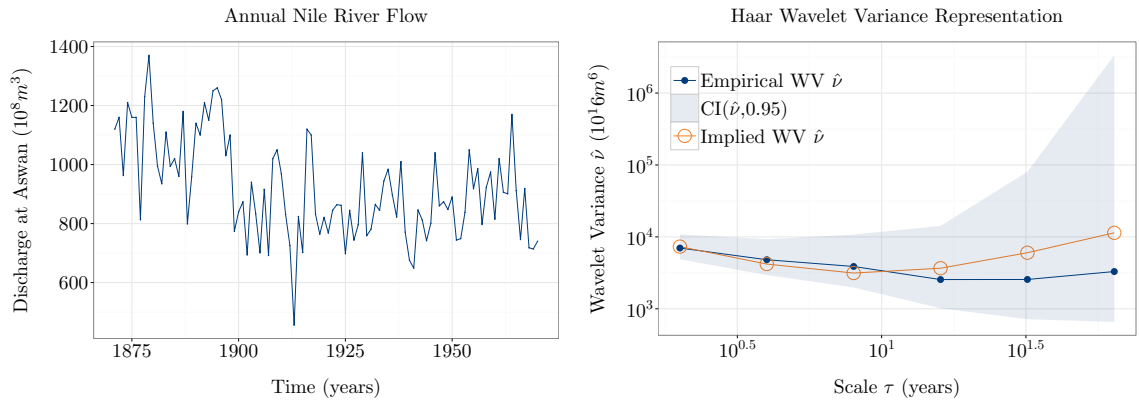

Figure 5: Left: Plot of Annual Nile river flow from 1871 - 1970. Right: WV implied by the selected local level model (orange line) and estimated WV (blue line) with confidence intervals (shaded area).

As highlighted at the end of Section 3, a first step to understanding if a robust analysis is needed, for example, is to compare the standard estimated WV with its robust counterpart and check if their confidence intervals overlap. If there are problems of robustness, these should be more evident at the first scales since the variability at the last scales often does not allow to detect a significant difference between the standard and robust WV. To illustrate this procedure, let us take another example given where we simulate from an `AR1()+WN()` model and then add noise coming from a normal random variable with variance equal to 100 to 1% of the observations. The code below shows how this is done and then computes the standard and robust WV on the resulting time series.

```
# Specify model
true.model = AR1(phi = .99, sigma2 = .01) + WN(sigma2 = 1)

# Generate time series
set.seed(213)
n = 1e3
sim.ts = gen.gts(true.model, n)

# Contaminate time series
cont.eps = 0.01
cont.num = sample(1:n,round(n*cont.eps))
sim.ts[cont.num] = sim.ts[cont.num] + rnorm(round(n*cont.eps),0,sqrt(100))

# Compute standard and robust WV
wv.classic = wvar(sim.ts)
wv.robust = wvar(sim.ts, robust = TRUE)

# Plot the robust WV
plot(sim.ts)
```

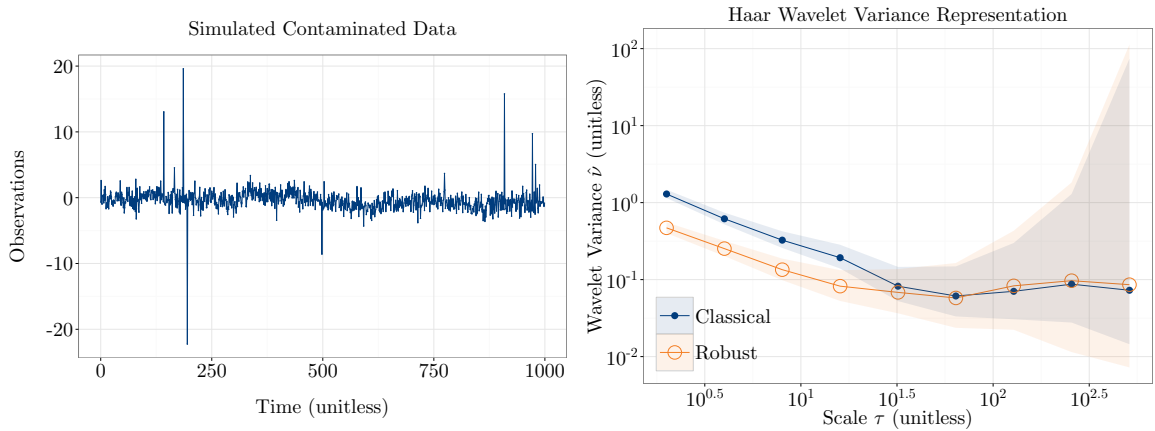

Figure 6: Left: Simulated time series with contaminated observations. Right: Comparison of resulting standard and robust WV.

```
compare.wvar(wv.classic, wv.robust, split = FALSE)

# Run robust estimation
rob.fit = gmwm(AR1()+WN(), sim.ts, robust = TRUE)

# Run inference
summary(rob.fit, inference = TRUE, bs.gof = TRUE)

Model Information:
        Estimates     CI Low    CI High           SE
AR1    0.98502549 0.96729698 0.99601635  0.008730068
SIGMA2 0.01433612 0.00629577 0.02184096  0.004725401
WN     0.95763144 0.86667884 1.03607348  0.051492314

Objective Function: 0.0285

Bootstrapped Goodness of Fit:
Test Statistic: 0.03
P-Value: 0.65  CI: (0.55, 0.74)

To replicate the results, use seed: 1337
```

The estimates of classic and robust WV, contained respectively in the objects `wv.classic` and `wv.robust`, can be compared using the function `compare.wvar()` seen earlier, using the option `split = F` to see the two estimates superposed. Figure 6 shows the result of this plotting function along with the plot of the simulated time series, highlighting how there appears to be contamination, confirmed by that fact that the two WV differ significantly across the first scales. Based on this analysis, a robust estimation procedure would appear necessary and for this purpose, as mentioned earlier, we use the function `gmwm()` with the option `robust = TRUE`. It must underlined that, when estimating the WV or model parameters robustly, it is possible to specify the desired level of efficiency of the robust estimator with respect to the standard WV and GMWM estimators. The default value is `eff = 0.6`, meaning that we want a low level of efficiency to obtain more robust estimates. However, the user can specify other values, usually between 0.5 and 1, the last value simply implying that the estimator corresponds to the standard estimator of WV (i.e. non-robust). It can be seen how the estimated parameters from the robust fit are close to the true simulation values despite the contamination in the time series.

Let us now consider a real data example given by the monthly precipitation series from 1907 to 1972 used in Hipel and McLeod (1994) which was obtained from DataMarket[6] and made available within the **datapkg** package[7]. In hydrology the study of water cycles is an extremely relevant topic and different models are used to easily interpret these cycles, such as the Environmental System Model of a watershed. The AR(1) model is among the candidates for the precipitation phase so let us estimate this model for this data (hereinafter `hydro`) using the code below.

```
# Check if there are robustness issues
wv = wvar(hydro)
rob.wv = wvar(hydro, robust = T)
compare.wvar(wv,rob.wv, split=F)

# Compare robust and non-robust fits
mle.fit = arima(hydro,order=c(1,0,0))
gmwm.fit = gmwm(AR1(), hydro)
rob.fit = gmwm(AR1(), hydro, robust=T)
```

The plot comparing the classic and robust WV (shown in Figure 7) indicates that there could be some contamination in the observations so we can compare the estimates of the MLE, GMWM and RGMWM to understand the influence of this contamination on the estimations. These are given in Table 2 where $\phi$ represents the AR(1) parameter and $\sigma^2$ represents the innovation variance. A significant difference can be observed between the non-robust estimators (i.e. MLE and GMWM) and the RGMWM suggesting that, even if the model is not the correct one, a robust estimation procedure is preferable for this dataset.

## 5. Computational Efficiency

The **gmwm** package has been developed and made available within the R programming language across Windows, Mac (OS X), and Linux with support for Solaris to arrive soon. Even though the framework's main residency is within R, the primary functions have been entirely implemented within C++ to obtain a high level of computational efficiency and to enable the ease of porting key functions to alternative computational frameworks. Moreover, the implemented functions use the Armadillo C++ Matrix Library API, which has a syntax that is very similar to Octave and R. As a result, the ability to extend the C++ library for users in both the engineering and statistical domains has a lower threshold.

---

[6] https://datamarket.com/data/set/22w1/mean-monthly-precipitation-1907-1972

[7] The **datapkg** package can be installed using `gmwm::install_datapkg()` and the source code is available at https://github.com/smac-group/datapkg

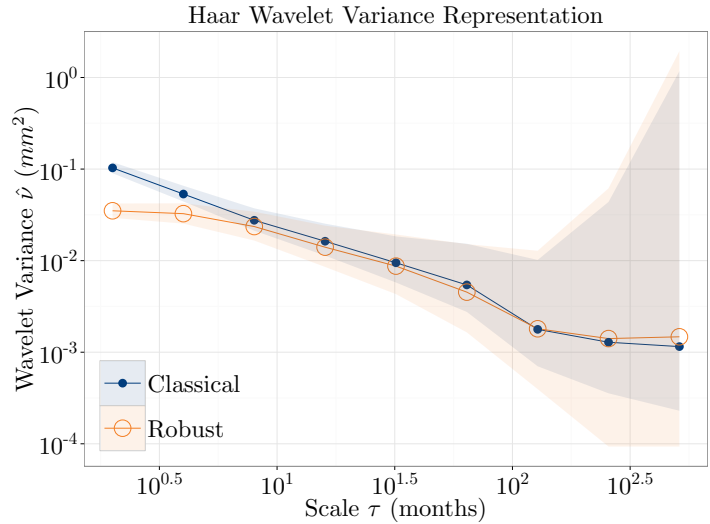

Figure 7: Comparison of the standard (blue line) and robust (orange line) estimated WV on the `hydro` data with shaded areas representing the respective confidence intervals.

|       | $\hat{\phi}$ | $\hat{\sigma}^2$ |
|-------|:---:|:---:|
| ML    | $6.843 \cdot 10^{-2}$ | $2.199 \cdot 10^{-1}$ |
| CI    | $[1.110 \cdot 10^{-3}, 1.212 \cdot 10^{-1}]$ | $[2.003 \cdot 10^{-1}, 2.3929 \cdot 10^{-1}]$ |
| GMWM  | $5.577 \cdot 10^{-2}$ | $2.179 \cdot 10^{-1}$ |
| CI    | $[-4.357 \cdot 10^{-3}, 1.153 \cdot 10^{-1}]$ | $[1.985 \cdot 10^{-1}, 2.365 \cdot 10^{-1}]$ |
| RGMWM | $4.049 \cdot 10^{-1}$ | $1.066 \cdot 10^{-1}$ |
| CI    | $[3.345 \cdot 10^{-1}, 4.662 \cdot 10^{-1}]$ | $[9.623 \cdot 10^{-2}, 1.175 \cdot 10^{-1}]$ |

Table 2: Estimated parameters for an `AR1()` model on the `hydro` dataset.

Part of the computational efficiency afforded to the package is due to the ability to interface between R and C++ with ease. The creation of this interface is through the considerable use of both Rcpp (Eddelbuettel, François, Allaire, Chambers, Bates, and Ushey 2011) and RcppArmadillo (Eddelbuettel and Sanderson 2014) to blend R data types (SEXP) with the GMWM C++ library. Particular attention has been placed on the use of this interface to avoid the trap of repeated calls from R into C++ thereby leading to excessive object copies during data type conversions and processing slowdowns. To reduce overall dependency on R and to ensure computational efficiency, some functions written in R were rewritten in C++ using Armadillo. The implementations are available in both the **gmwm** package and in the "R to Armadillo" repository[8]. The rewritten implementations typically yield an increase in computational speed between 1.5 and 4 times faster and also serve as well documented examples of porting R code into C++ using Rcpp.

---

[8]R to Armadillo source is available at: https://github.com/coatless/r-to-armadillo

Having discussed the software framework, in this section we briefly underline the computationally efficiency of the **gmwm** package in estimating the WV and the models mentioned in Sections 4.1 and 4.2. To provide a first example of the computational speed of the functions in the **gmwm** package, let us compare the speed at which the WV is computed using the `wave.variance()` function within the **waveslim** package and the new `wvar()` function in the **gmwm** package, including the robust option. Table 3 collects the results of one hundred benchmark tests using the **rbenchmark** package.

|          | Sample size | | | | |
|---------:|:------:|:------:|:------:|:------:|:------:|
| Package  | 100    | 1,000  | 10,000 | 100,000 | 1,000,000 |
| **waveslim** | 0.0004 | 0.0013 | 0.0086 | 0.0823 | 0.7432 |
| **gmwm** | 0.0001 | 0.0001 | 0.0011 | 0.0223 | 0.3575 |
| **gmwm** (robust) | 0.0069 | 0.0125 | 0.0530 | 0.6531 | 9.6700 |

Table 3: Benchmarks of the WV computation in seconds.

It can be observed how the new function `wvar()` considerably increases the speed at which the WV is computed compared to the already existing `wave.variance()` function, going at least twice as faster than the latter which is of considerable importance when carrying out analyses on large samples. On the other hand, the robust estimation is visibly slower, as would be expected, preserving however reasonable computational times even in large samples.

Let us now consider the computational times when estimating two models:

- `ARMA(3,1)`: autoregressive moving average model;

- `2*AR1()+WN()`: a latent model made by the sum of two AR(1) processes and a white noise.

We estimate the parameters of these models 10 times with the following estimators:

- MLE;

- GMWM;

- Robust GMWM (RGMWM) presented further on in Section 4.3;

- Indirect inference based on an auxiliary AR(7) model using robust estimation proposed by Kunsch (1984) (KUNSCH).

For the `ARMA(3,1)` model, only the robust estimators were compared (i.e. RGMWM and KUNSCH) while all of them were compared for the latent model. Therefore each GMWM estimator is compared with an alternative that is available (or easily implemented) in R. In all cases, the optimization procedures used the true parameter values as starting values. Figure 8 shows the median computation times for these estimators on the two models for sample sizes going from 100 to 10 million, with a roof for computation placed at 6 hours.

It can be seen how the GMWM and RGMWM estimators have extremely low computational times compared to the other estimators in the two cases. Indeed, the computational times for the KUNSCH estimator are beyond the chosen maximum time limit for sample sizes larger than 100,000 while the RGMWM remains under 1 hour even for samples of size 10 million.

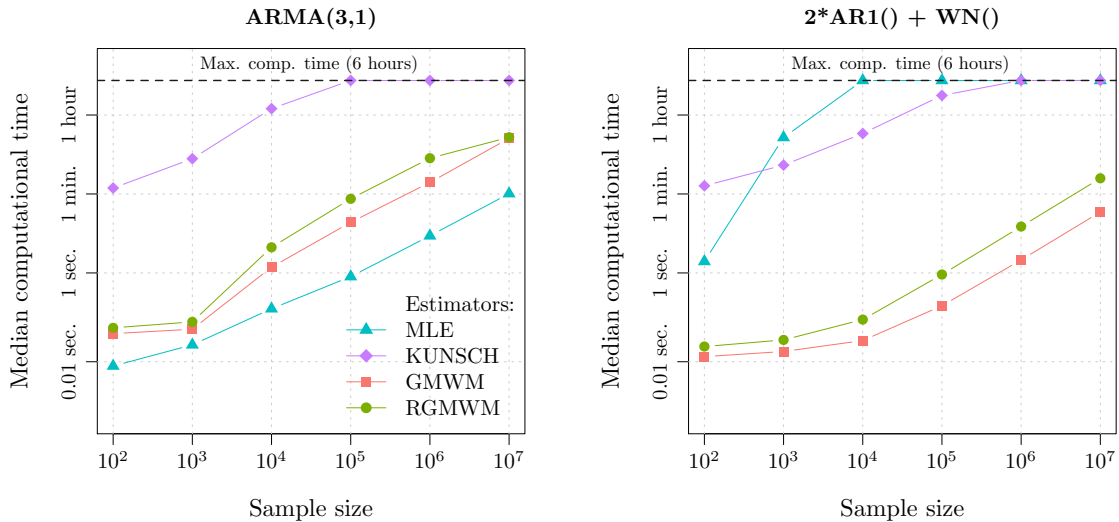

Figure 8: Median estimation times of the methods on different sample sizes.

Similar conclusions can be made for the latent model where the GMWM estimators are clearly faster than the other two methods in all cases. The RGMWM is slower than the GMWM as would be expected but it is able to estimate the latent model in a little over two minutes for sample sizes of 10 million. These results show that not only the **gmwm** package is able to consistently estimate a wide range of time series models, but it is also able to do so robustly and in an extremely computationally efficient way even for very large sample sizes.

## 6. Conclusion

The implementation of the GMWM framework within the **gmwm** package represents a first step in the direction of a platform which delivers a wide set of computationally efficient tools for (robust) time series model estimation and inference. Aside from the many advantages highlighted in this paper, the main contributions that this package delivers are the ease with which linear state-space models can be estimated and the tools for the robust estimation of a wide range of time series models. These two aspects considerably reduce the barriers to which users were subject to due to the unavailability, complexity and/or high computational demand of existing statistical software. In addition, the nature of this estimator and the idea of latent model structures allows it to be extended to different settings.

**Affiliation:**

Stéphane Guerrier
Assistant Professor
Department of Statistics
University of Illinois at Urbana-Champaign
725 S. Wright St.
Champaign, IL 61820 E-mail: stephane@illinois.edu
URL: http://publish.illinois.edu/stephane